\documentclass[twocolumn, tighten]{aastex701}
\usepackage{amsmath}
\usepackage{amssymb}
\usepackage{natbib}
\usepackage{xcolor}
\usepackage{hyperref}

\usepackage{newtxtext,newtxmath}
\usepackage[T1]{fontenc}
\usepackage{graphicx}
\usepackage{caption}
\usepackage{tabularx}
\usepackage{pifont}

\newcommand{\ovii}{O{\small VII}}

\begin{document}

\title[CGM mass with X-ray resonant scattering]{Inferring the mass of the circumgalactic medium using X-ray resonant scattering}
\author[0000-0003-4983-0462]{Nhut Truong}
\affiliation{Center for Space Sciences and Technology, University of Maryland, Baltimore County, 1000 Hilltop Circle, Baltimore, MD 21250, USA}
\email{ntruong@umbc.edu}
\affiliation{NASA Goddard Space Flight Center, Greenbelt, MD 20771, USA}
\author[0000-0003-0144-4052]{Maxim Markevitch}
\affiliation{NASA Goddard Space Flight Center, Greenbelt, MD 20771, USA}
\email{maxim.markevitch@nasa.gov}
\author[0000-0001-8421-5890]{Dylan Nelson}
\affiliation{Universit\"{a}t Heidelberg, Zentrum f\"{u}r Astronomie, ITA, Albert-Ueberle-Str. 2, 69120 Heidelberg, Germany}
\email{dnelson@uni-heidelberg.de}
\author[0000-0002-0885-8090]{Chris Byrohl}
\affiliation{Universit\"{a}t Heidelberg, Zentrum f\"{u}r Astronomie, ITA, Albert-Ueberle-Str. 2, 69120 Heidelberg, Germany}
\affiliation{Kavli Institute for Particle Astrophysics \& Cosmology (KIPAC), Stanford University, Stanford, CA 94305, USA}
 \email{chris.byrohl@uni-heidelberg.de}

 \correspondingauthor{Nhut Truong} 
 \email{ntruong@umbc.edu; nhut.truong@nasa.gov}

\begin{abstract}
The circumgalactic medium (CGM) regulates galaxy growth and retains the imprint of feedback from supernovae and supermassive black holes. However, the bulk of the hot CGM produces little X-ray emission and is challenging to study with X-ray telescopes.
We propose a novel method for evaluating the CGM mass using resonant scattering of the helium-like oxygen (\ovii) resonant line at $E=574$ eV. 
In a spherically symmetric and static CGM halo with a sharp central X-ray peak, the number of \ovii\ ions within an outer radial shell can be calculated from the ratio of the two directly observable quantities: the \ovii\ flux from the bright inner region and the scattered \ovii\ flux from the shell (where the scattered flux can be much higher than the intrinsic emission). 
To evaluate the accuracy of this geometric estimate for realistic galaxies --- with satellites, asymmetries, and gas velocities --- we use a sample of galaxies from the TNG50 cosmological simulation. 
We find that, when the most irregular systems are excluded based on their X-ray observables, we accurately predict the \ovii\ mass in the outer halo (e.g., in an $r=R_{\rm 500c}-R_{\rm 200c}$ shell) from the ratio of the fluxes in the corresponding annulus and the central peak region ($r<0.2R_{\rm 500c}$), with only a 10\% bias and an rms scatter of $\sim 0.2$ dex. 
As \ovii\ mass strongly correlates with the total oxygen and gas mass, this direct \ovii-counting method enables indirect estimates of those quantities by future X-ray microcalorimeter missions, such as {\em NewAthena}\/ and {\em HUBS}.

\end{abstract}


\section{Introduction}

According to the standard model of structure formation, small primordial density perturbations collapse under their own gravity and form galactic halos that consist of dark matter (DM) and gas. Some of the gas rapidly cools through radiation, becomes thermally unstable and condenses into stars and potentially forms a supermassive massive black hole (SMBH) at the halo center. The remaining gas is shock heated to approximately the halo's virial temperature, forming halos of hot ($T\sim10^{6}K$), volume-filling circumgalctic medium (CGM) around massive galaxies \citep{white.rees.1979, white.frenk.1991}. The CGM plays a crucial role in governing the formation and evolution of galaxies, serving as a gas reservoir that fuels the star formation. It also acts as the interface between the inflow of pristine intergalactic gas and the outflows driven by galactic ``feedback'' processes, such as supernovae and jets produced by accretion onto SMBH (see \citealt{werner.etal.2019} for a review).

The CGM encodes signatures of those galactic feedback processes. Though recognized as an important actor in galaxy formation, the hot CGM remains largely elusive. UV observations with the Hubble Space Telescope detect and characterize the warm ($T\sim 10^5$\,K) phase of the CGM. These observations use bright background quasars to measure absorption by the CGM along individual sightlines \citep{tumlinson.etal.2017}, with sampling of the CGM atmospheres that is unavoidably sparse. X-ray observations with {\em XMM-Newton}\/ and {\em Chandra}\/ detect extended X-ray emission from the hot CGM in its inner regions ($r\lesssim 50$ kpc $\sim 0.2 R_{\rm 200c}$%
\footnote{$R_{Xc}$ is the radius within which the average matter density of the halo is $X$\/ times the Universe's critical density.}%
) in the most massive galaxies \citep{mineo.etal.2012,bogdan.etal.2017,li.etal.2017}. The bulk of the CGM resides beyond those radii, but is too faint to be detectable with current X-ray instruments in individual halos. Recently, stacking of X-ray data from tens of thousands of galaxies using the eROSITA All-Sky Survey \citep{comparat.etal.2022,chadayammuri.etal.2022, zhang.etal.2024a,zhang.etal.2024b, zhang.etal.2024c} led to a tentative detection of extended emission for samples of galaxies with masses of the Milky Way and above. However, the interpretation of these detections remains complex due to the unknown contribution to the stacked X-ray profiles from individual galaxies. The stacked signal could be dominated by a handful of exceptionally X-ray bright galaxies, e.g., those residing in crowded environments, such as galaxy groups \citep{zhang.etal.2024c}.

The hot CGM can also be detected via the Sunyaev-Zel'dovich (SZ) effect, which is proportional to the plasma electron pressure \citep{sunyaev.zeldovich.1969}. Studies using data from Planck and Atacama Cosmology Telescope reported SZ detections of the hot CGM in stacked samples of galaxies at $z<0.1$ \citep{planck13, greco.etal.2015, lim.etal.2018,das.etal.2023}. The stacked SZ signal can derive average CGM pressure profiles for a galaxy population, but, similarly to X-ray stacking studies, it can be biased by outliers.

Probing the extended CGM in individual galaxies spanning a range of masses and types is essential to form a full picture of galaxy formation.
It is especially important to probe galaxies above and below the pivotal Milky Way mass scale ($M\sim 10^{12}$ $M_\odot$), where simulations suggest that the physics of the galaxy-CGM interplay changes qualitatively \citep[e.g.,][]{dekel.etal.2006,faucher.etal.2011,nelson.etal.2018}.
One way to probe the extended CGM in individual galaxies is via X-ray line emission. The CGM has temperatures in the $10^{5.5-7}$K range, where the X-ray luminosity is dominated by emission lines from abundant ions such as C{\small V}, C{\small VI}, O{\small VII}, O{\small VIII}, and Fe{\small XVII}\  \citep{truong.etal.2023, schellenberger.etal.2024}.%

This faint redshifted line signal from the galaxies could be separated from the much brighter Milky Way foreground of the $z=0$ emission lines using the future X-ray microcalorimeter arrays, such as {\em NewAthena}\/ and {\em HUBS}\/ \citep{barret.etal.2016, bregman.etal.2023}.

Mapping the plasma thermal emission in the X-ray lines will reveal the presence of the CGM and probe its spatial distribution. However, it would still not allow us to quantify the properties of the CGM such as its density (and therefore mass and metallicity), because the X-ray line emission alone (without the X-ray continuum) is not sufficient to break the degeneracy between density, temperature and metallicity. As pointed out in \cite{kraft.etal.2022} and \cite{schellenberger.etal.2024}, the CGM X-ray continuum is far below the level of the Milky Way foreground, which is 
a limitation that cannot be overcome with more powerful X-ray telescopes. Even if it were accessible, the X-ray emission is proportional to density squared and the assumptions of symmetry and uniformity are required to derive the CGM mass. Therefore, it is crucial to find new ways to infer the  physical properties of the CGM.

Most of the CGM is optically thin for the X-ray emission, with the exception of resonant transitions \citep{gilfanov.etal.1987,churazov.etal.2010} of the most abundant ions, and in particular, the Helium-like Oxygen {\small VII}. The brightest \ovii\ emission line complex is the He$\alpha$ triplet that includes the resonant ({\em r}), intercombination ({\em i}), and forbidden ({\em f}) lines at rest-frame $E=574,\,569,\,561$ eV, respectively; hereafter we denote the resonant component as \ovii{\em r}. The scattering optical depth, $\tau$, in the \ovii{\em r}\/ line can be significant, although still $\ll$1 over most of the CGM volume. This results in a small (factor $1-\tau$) suppression of the \ovii{\em r}\/ flux that reaches the observer from the bright central regions and a commensurate absolute flux enhancement at larger radii. Because the intrinsic (thermal plasma) \ovii{\em r}\/ surface brightness in the outskirts is much lower than at the center, this enhancement can dominate the \ovii{\em r}\/ flux, sometimes exceeding the intrinsic brightness by over an order of magnitude \citep{nelson.etal.2023}. \cite{churazov.2001} and \cite{khabibullin.churazov.2019} pointed out that the X-ray resonant line emission from the warm-hot intergalactic medium (WHIM) can be dominated by the scattered emission from the point sources that constitute the cosmic X-ray background (CXB), greatly enhancing the detectability of the WHIM emission. The CGM situation is similar, except there is always a much brighter, nearby source illuminating the CGM outskirts, which is the central, bright region of the same CGM halo.

In this paper, we go beyond the mere detection of low-surface-brightess CGM regions using \ovii\ scattering \citep{nelson.etal.2023} and propose a method to count the number, and hence the mass, of \ovii\ ions in a CGM halo that is based on scattering. We test the accuracy of this method using galaxies from the TNG50 suite of cosmological magnetohydrodynamical simulations, and propose observational criteria to select a sample of real galaxies for which this method should produce an accurate mass estimate.

The paper is arranged as follows. In Section \ref{sec:method}, we describe the method idea, the sample of galaxies from TNG50, and radiative transfer (RT) treatment. Sections \ref{sec:3}-\ref{sec:systematics} describe the correlation between the \ovii\ mass of the simulated galaxies and the observables at the basis of our method, examine systematic effects that introduce scatter, and discuss strategies to mitigate these effects in real observations. In Section \ref{sec:theory_vs_sim}, we compare the theoretically derived scattering scaling relation with results from the simulation. Finally, Section \ref{sec:mass_scaling} discusses empirical scaling relations between \ovii\ mass and other CGM mass components. 

%%%%%%%%%%%%%%%%%%
\begin{figure}
    \centering
	\includegraphics[width=0.4\textwidth]{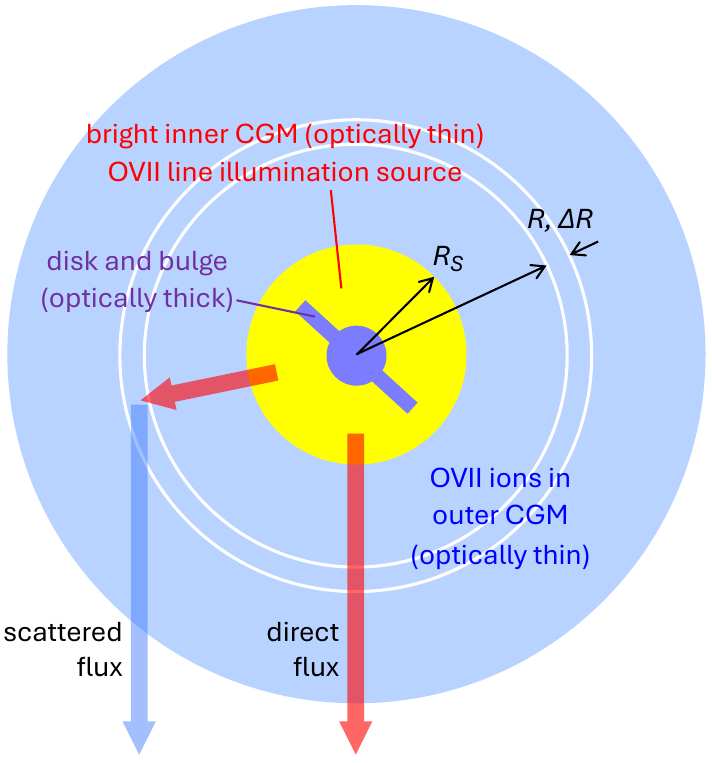}
    \vspace*{1mm}
    \caption{Resonant scattering can be used to deduce the mass of \ovii\ ions in a CGM halo. Most \ovii{\em r}\/ photons are emitted within the inner CGM. These photons scatter off \ovii\ ions in the outer CGM, where the intrinsic \ovii{\em r}\/ emission is negligible in comparison. The observer sees both the direct \ovii{\em r}\/ emission from the central (source) region and photons from the same source scattered off the \ovii\ ions in the outer region. The ratio of these fluxes gives the total number of \ovii\ ions in the outer region. Both regions are reasonably optically thin (except for small fractions occupied by the disk and bulge), which means that ions throughout the outskirts see the same flux from the central region as does the observer.}
    \label{fig:O7_scattering}
    \vspace*{2mm}
\end{figure}
%%%%%%%%%%%%%%%%%%%

%%%%%%%%%%%%%%%%%%%%%%%%%%%%%%%%%%%%%%%
\section{HOW TO COUNT \ovii\ IONS IN A CGM HALO}
\label{sec:method}

\subsection{The idealized case}
\label{sec:theory}

The surface brightness distribution of the \ovii{\em r}\/ emission over a CGM halo can be used to derive the total number of \ovii\ ions in the outer confines of the halo, where most of the CGM mass resides. The concept is shown in Figure \ref{fig:O7_scattering}. We first consider an idealized spherically symmetric, static, optically thin halo with a sharply centrally peaked brightness profile of the plasma thermal X-ray emission, as observed in real galaxies. Most \ovii{\em r}\/ photons are emitted in the central peak, within a small galactocentric radius $R_{\rm S}\ll R_{\rm 200c}$. These photons scatter off the \ovii\ ions in the outer CGM, and some of them reach the observer. We assume that scattering occurs isotropically with an averaged cross-section $\langle\sigma_{\rm scat}\rangle$. The optical depth through the wall of a uniform shell with radius $R$\/ and thickness $\Delta R\ll R$ is given by
\begin{equation}
    \tau = \langle\sigma_{\rm scat}\rangle\, n_{\rm OVII}\, \Delta R = \frac{\langle\sigma_{\rm scat}\rangle N_{\rm OVII}}{4 \pi R^2}
\label{eqn3}
\end{equation}
where $n_{\rm OVII}$ is the number density and $N_{\rm OVII}$ is the total number of \ovii\ ions in the shell. If $\tau\ll 1$, the luminosity of the scattered \ovii{\em r}\/ emission from the shell can be approximated as
\begin{equation}
L_{\rm shell} = (1-e^{-\tau})L_{\rm source}\approx L_{\rm source}\, \tau,
\label{eq:l}
\end{equation}
where $L_{\rm source}$ is the \ovii{\em r}\/ luminosity of the source region; then
\begin{equation}
N_{\rm OVII}= 
\frac{4\pi R^2}{\langle\sigma_{\rm scat}\rangle} \frac{L_{\rm shell}}{L_{\rm source}}.
\label{eq:novii}
\end{equation}
Because the CGM within the central peak region can be considered optically thin too (for the most part --- except the galactic disk and bulge at the center of the source region), the emission from the source region is isotropic. Thus, 
\begin{equation}
\frac{L_{\rm shell}}{L_{\rm source}}=\frac{F_{\rm shell}}{F_{\rm source}},
\label{eq:f}
\end{equation}
where $F_{\rm shell}$ and $F_{\rm source}$ are \ovii{\em r}\/ fluxes from the respective regions seen by a distant observer. Given other uncertainties that will be discussed below, here we disregard scattering on the path from the source through the outer CGM layers: $\tau (R>R_S)\ll 1$. This flux ratio can be used to evaluate the number of \ovii\ ions in each thin shell throughout the outer CGM halo.

Equation (\ref{eq:novii}) is applicable only in the thin-shell limit, i.e. $\Delta R\ll R$. In the more general situation, the \ovii\ ions number can be expressed as
\begin{equation}
 N_{\rm OVII}= \frac{V_{\rm shell}(R, \Delta R)}{\Delta R}\langle\sigma_{\rm scat}\rangle^{-1}\bigg(\frac{F_{\rm shell}}{F_{\rm source}}\bigg)
\label{eq:movii}
\end{equation} 
where $V_{\rm shell}$ is the shell's volume. To convert the number of \ovii\ ions into an \ovii\ mass, we multiply both sides of the above equation by the oxygen atomic mass. Throughout the remainder of the paper, we present results in term of the total \ovii\ mass ($M_{\rm OVII}$) contained within the shell. 

At large $R\gtrsim R_{500c}$, the intrinsic thermal emission in \ovii{\em r}\/ is negligible compared to the scattered emission \citep{nelson.etal.2023}. For the radii where it is not negligibly small, one would be able to separate the two using the fact that scattering produces only the \ovii{\em r}\/ signal, while the intrinsic emission is the triplet with the ratio of the {\em f, i, r}\/ components that is known to a sufficient accuracy.

The cross-section $\sigma_{\rm scat}$ depends on the \ovii{\em r}\/ spectral line profile. In a static CGM with no thermal or turbulent gas motions, the line profile is Lorentzian with a width set by the \ovii{\em r}\/ radiative (natural) width $\Gamma_{\rm rad}=\frac{h}{2\pi}A\approx2\times10^{-3}$ eV, where $A\approx3.32\times10^{12}\ {\rm s}^{-1}$ is the Einstein coefficient for the \ovii\ resonant transition \citep{rybicki.lightman.1979,verner.verner.ferland.1996}, and $h$ is the Planck constant.

In a realistic CGM, the line profile is broadened by gas motions. At a typical temperature of $\sim10^6$ K, the thermal Doppler width%
\footnote{Throughout this paper, we characterize line broadening by Doppler width $W_D$, which is related to the Gaussian profile parameter $\sigma$ as $W_{D}=\sqrt{2}\sigma$.}
is $\sim0.06$ eV, while turbulent motions with a typical velocity dispersion of $\sim100$ km/s contribute a Doppler broadening of $\sim0.3$ eV---both far exceed the radiative width of $\sim0.002$ eV. The line width is therefore determined by Doppler broadening. 

Assuming \ovii\ ions in the shell have a Gaussian velocity distribution, which produces a Gaussian line profile centered at $E_{\rm shell}$ with Doppler width $W_{D,{\rm shell}}$, the scattering cross-section for an incoming photon with energy $E$ is given by
\begin{equation}
    \sigma_{\rm scat} (E)= \frac{\sqrt{\pi} h r_e c f}{W_{D, {\rm shell}}}\exp\bigg[{-\frac{\big(E-E_{\rm shell}\big)^2}{W_{D, {\rm shell}}^2}}\bigg], 
\label{eqn5}    
\end{equation}
where $r_e$ is the classical electron radius, $c$\/ is the speed of light, and $f$\/ is the oscillator strength. The Doppler broadening width, $W_{D, {\rm shell}}$, is
\begin{equation}
    W_{D, {\rm shell}}  = E_{\rm shell} \bigg[\frac{2k_B T}{A_O m_p c^2}+\frac{2v_{\rm turb}^2}{c^2}\bigg]^{1/2}, 
\label{eqn6}
\end{equation}
where $k_B$ is the Boltzmann constant, $T$ is CGM temperature in the shell, $A_{O}$\/ is the oxygen atomic mass, $m_p$ is the proton mass, and $v_{\rm turb}$ is 1D turbulent velocity dispersion. Assuming that scattering photons originate from a source whose energies also follow a Gaussian distribution with mean energy $E_{\rm source}$ and Doppler width $W_{D, {\rm source}}$
\begin{equation}
f_{\rm source} (E) = \frac{1}{\sqrt{\pi}W_{D,{\rm source}}}\exp\bigg[-\frac{\big(E-E_{\rm source}\big)^2}{W_{D, {\rm source}}^2}\bigg],
\end{equation}
the averaged cross-section is given by
\begin{eqnarray}
&\langle \sigma_{\rm scat} \rangle&= \int_{-\infty}^{\infty}dE\ \sigma_{\rm scat}(E)\times f_{\rm source}(E) \nonumber \\
&&= \frac{\sqrt{\pi} h r_e c f}{\sqrt{W_{D, {\rm shell}}^2+W_{D,{\rm source}}^2}}\exp\bigg[{-\frac{\big(E_{\rm source}-E_{\rm shell}\big)^2}{W_{D, {\rm shell}}^2+W_{D, {\rm source}}^2}}\bigg].\nonumber \\
\label{eqn:sigma_scat}
\end{eqnarray}
In general, the mean energies $E_{\rm source}$ and $E_{\rm shell}$ may differ from the rest-frame line center energy $E_0=574$ eV due to bulk motion of the emitting and scattering gas; however, only their difference enters the averaged cross-section. Resonant scattering is effective when the line center offset between the source photons and the scattering ions is small compared to the combined Doppler width, and is suppressed when the relative velocity is large.

The above idealized derivation includes several simplifying assumptions, the most important one being that the large-scale CGM motions are small and there is no significant Doppler mismatch between the source and the outskirts that would cause $\langle \sigma_{\rm scat} \rangle$ go to zero (eq.\ \ref{eqn:sigma_scat}) and make $N_{\rm OVII}$ in eq.\ (\ref{eq:novii}) unconstrained. Other possible sources of error are asymmetries in the central bright region, contamination by intrinsic emission from subhalos in (or projected onto) the outer shell, and the shell's non-uniformity. Below we evaluate these complications for realistic halos and devise ways to minimize them observationally.

\subsection{Realistic GCM halos from the TNG50 galaxy sample}

To evaluate the accuracy of the idealized scaling relation (\ref{eq:movii}) for realistic CGM halos, and to identify sample selection criteria that the future X-ray observer might apply to maximize the accuracy of the estimate, we use a sample of galaxies from the TNG50 cosmological magnetohydrodynamical simulation (\citealt{pillepich.etal.2019,nelson.etal.2019b}). TNG50 is the third simulation suite of the IllustrisTNG project (TNG hereafter; \citealt{marinacci.etal.2018, naiman.etal.2018, springel.etal.2018, nelson.etal.2018, pillepich.etal.2018b}). Of the TNG simulations, TNG50 features the highest numerical mass resolution, $ m_{\rm baryon}\sim8\times10^{4}M_\odot$, within a simulated box of ${\rm \sim50\ cMpc}$ per side. All the TNG simulations are performed with the AREPO moving-mesh code (\citealt{springel.2010}) and adopt cosmological parameters consistent with the \citealt{planck.2016}: $\Omega_{\rm m,0}=0.6911$, $ \Omega_{\rm m,0}=0.3089$, $\Omega_{\rm b,0}=0.0486$, $\sigma_8=0.8159$, $ n_s=0.9667$, and $h=0.6774$.

The TNG simulations employ a galaxy formation model that incorporates a variety of astrophysical processes crucial for galaxy formation (\citealt{weinberger.etal.2017, pillepich.etal.2018}). These include radiative cooling, from primordial and metal-line processes, plus heating from the UV background radiation. It also models star formation, stellar evolution, and chemical enrichment via supernovae type Ia, II, and AGB stars, allowing the simulation to track the evolution of nine individual elements: H, He, C, N, O, Ne, Mg, Si, and Fe. In addition, the TNG model realizes feedback arising from supernovae (stellar feedback) and active galactic nuclei (AGN). SMBHs are seeded into massive halos and grow through accretion or mergers with other SMBHs, while AGN feedback adopts a two-mode scenario, thermal or kinetic, depending on accretion rate \citep{weinberger.etal.2017}.

In this study, we adopt the same galaxy sample used in \cite{nelson.etal.2023}. This sample consists of 249 galaxies drawn from a parent sample of 479 central galaxies in TNG50 at $z=0$ with stellar mass in the range $10^{10.0}<M_*/M_\odot<10^{11.0}$. From the parent sample, galaxies are randomly subsampled up to 30 per 0.1 dex bin of $M_*$. 

%%%%%%%%%%%%%%%
\begin{figure*}[ht!]
    \centering
	\includegraphics[width=0.97\textwidth]{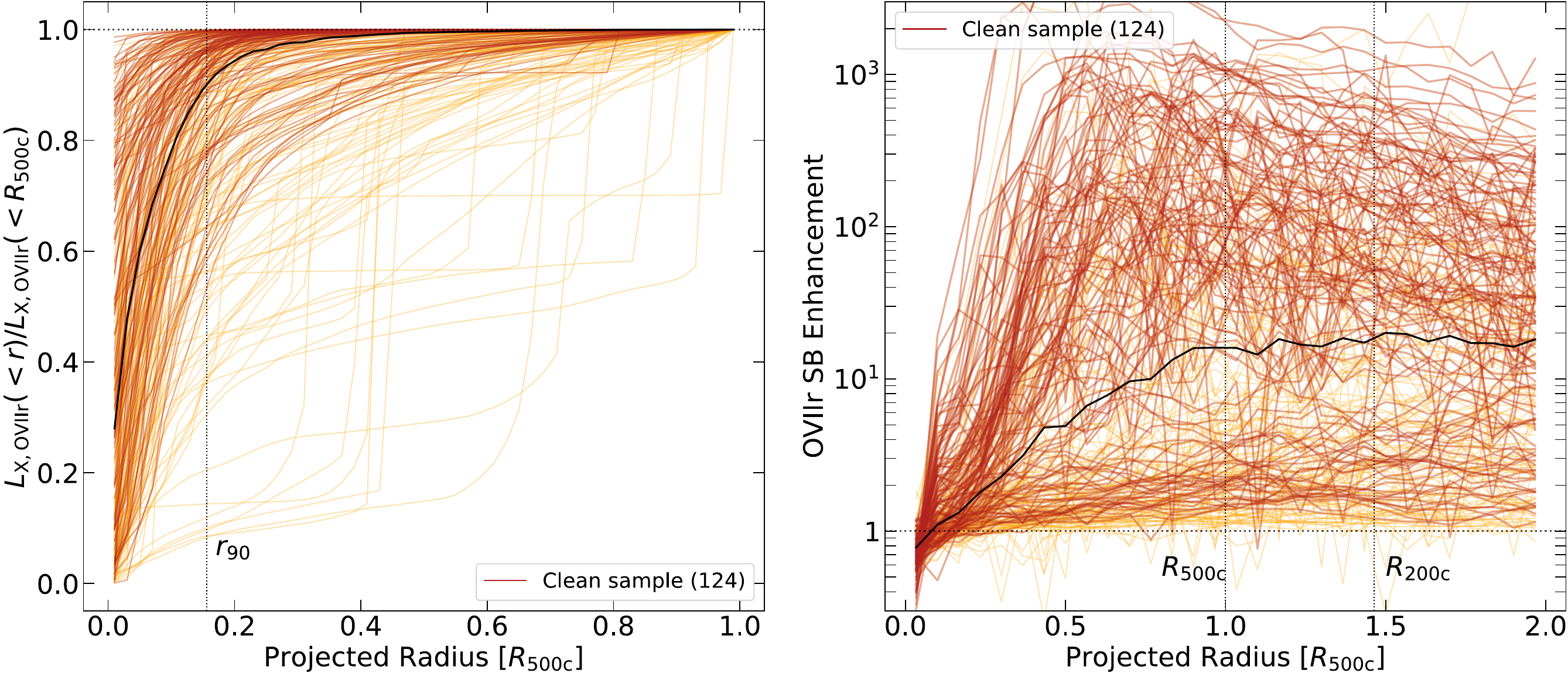}
    \vspace*{1mm}
    \caption{Definition of the inner (source) and outer CGM regions based on \ovii{\em r}\/ emission profiles and scattering enhancement. {\it Left:} The \ovii{\em r}\/ cumulative intrinsic (i.e., not considering resonant scattering) emission profiles, normalized to the total emission within $R_{\rm 500c}$, are shown as a function of projected radius. Individual profiles are shown by thin lines, and the thick black line shows the sample median. The vertical dotted line marks the median radius within which $90\%$ of ${L_{\rm X}(<R_{\rm 500c})}$ originates (denoted as $r_{\rm 90}$). Most of the \ovii{\em r}\/ emission originates within ${\sim 0.2R_{\rm 500c}}$. {\it Right:}\/ The \ovii{\em r}\/ surface brightness enhancement factor due to the resonant scattering. The vertical dotted lines indicate the sample median values of $R_{\rm 500c}$ and $R_{\rm 200c}$. On average, the scattering-driven enhancement in the \ovii{\em r}\/ surface brightness increases toward the galaxy outskirts and plateaus at $r>R_{\rm  500c}$. The color coding is the same for both panels: the clean subsample (red curves), which excludes irregular galaxies identified based on X-ray observables (yellow curves), is defined in Section \ref{sec:systematics}.} 
    \label{fig:boundary}
    \vspace*{3mm}
\end{figure*}
%%%%%%%%%%%%%%%%

\subsection{Radiative transfer of \ovii{\em r}\/ photons}

The \ovii{\em r}\/ emission and scattering is modeled in postprocessing, based on the simulated thermodynamic and kinematic properties of the CGM in individual galaxies. In the following, we briefly summarize the radiative transfer treatment of the selected galaxies, which is described in detail in \cite{nelson.etal.2023}.

\subsubsection{Modelling of \ovii\ ion abundances and \ovii{\it r}\/ line emission}

The CGM ion abundances and \ovii{\em r}\/ line emission are calculated using the CLOUDY package (\citealt{ferland.etal.2017}, version 17.00). We use a CLOUDY model that includes both collisional and photoionization processes under the condition of ionization equilibrium in the presence of UV and X-ray backgrounds (\citealt{faucher-Giguere.etal.2009}). From this CLOUDY model, we generate tables that tabulates emissivity and abundance fraction as functions of density, temperature, redshift, and metallicity. These tables are then used to interpolate the emissivisity and ion abundance for each gas cell based on its thermodynamics and metal content. 

The hot interstellar medium (ISM) is represented in the TNG simulations with a two-phase model (\citealt{springel.hernquist.2003}). We calculate the mass fraction, and the respective density and temperature of the hot ISM phase via density-dependent relations provided by the two-phase model. These values are then input to the CLOUDY model to compute \ovii{\em r}\/ emission from the hot ISM. As noted in \cite{nelson.etal.2023}, this approach provides an approximation to account for strong \ovii{\em r}\/ emission from the hot ISM, but requires a free scaling parameter $b$\/ that can adjust the \ovii{\em r}\/ emission from the multiphase ISM. This parameter also allows us to gauge the impact of a bright source within the central region of the galaxy such as the galaxy itself, or a luminous AGN. We adopt a fiducial value of $b=10^{-4}$. For a more thorough discussion on the use of this boost parameter, we refer to \cite{nelson.etal.2023}.

This model for \ovii{\em r}\/ emission and \ovii\ abundance is based on a number of assumptions. First, (i) it is limited by the numerical resolution of the TNG50 simulation, with no emission modelled below the physical resolved scales of $\sim100$ pc. Second, (ii) continuum emission in the \ovii{\em r}\/ line energy range from the stellar component is not included. Third, (iii) the effects of radiation from stellar or AGN sources on the ionization state of the CGM gas (e.g., \citealt{suresh.etal.2017, oppenheimer.etal.2018}), are not accounted for. We stress that in real galaxies, the central \ovii{\em r}\/ flux that enters our mass estimate will be directly observed, so the above modeling uncertainties are unimportant for the purposes of testing our method, as long as the simulated fluxes are qualitatively realistic.

\subsubsection{Radiative transfer treatment}
\label{sec:RTsim}

The resonant scattering process of \ovii{\em r}\/ line is modeled with Monte Carlo radiative transfer \citep[MCRT:][]{byrohl.etal.2021}, extending previous work on Ly$\alpha$ to X-ray resonant lines with the new THOR code\footnote{\hyperref[https://thor-rt.org/]{https://thor-rt.org/}} \citep{byrohl.2025}. The rest-frame wavelength of the \ovii{\em r}\/ line is $\lambda=21.602\Angstrom$, with oscillator strength $f=0.696$ and emission coefficient ${\rm A=3.32\times10^{12}\ s^{-1}}$ (\citealt{verner.verner.ferland.1996}).

In the RT calculations, \ovii{\em r}\/ photons are emitted at wavelengths determined by the velocity and temperature of the emitting gas cell. For the latter, we assume a thermal Gaussian distribution. Turbulent motions at scales below the numerical resolution limit ($\sim100$ pc) are not accounted for. 

Photons are emitted and scattered isotropically in the RT simulation. This is a simplification: the phase function for the \ovii\ resonant transition is slightly anisotropic and should depend on the scattering angle $\theta$ as $p(\theta)\propto (1+\cos^2{\theta})$ \citep{hamilton.1947, chandrasekhar.1950}. Random gas motions in the CGM would partially isotropize this angular distribution. Because the simulation has been performed assuming isotropic scattering, we must follow the same assumption here for consistency; as long as our estimate and the simulation are self-consistent, this does not affect the evaluation of the accuracy of our method. We discuss this further in \S\ref{sec:summary}.

The RT treatment accounts only for resonant scattering of \ovii{\em r}\/ photons, preserving the number of photons and leaving the underlying gas state unchanged.
Upon emission and each subsequent scattering, we spawn ``peeling-off contributions'' \citep{YusefZadeh84} computing the average luminosity contribution to reach the observer along a particular line of sight. Saving these contributions, as well as the original emitted contributions, allows us to compute mock observations with and without scattering effects.

%\section{Mass scaling relation induced by resonant scattering}
\section{Application to simulated galaxies}
\label{sec:3}
%%%%%%%%%%%%%%%%%%%%%%%%%%%%%%%%%%%%%%%
\begin{figure*}[ht!]
    \centering
    \includegraphics[width=0.72\textwidth]{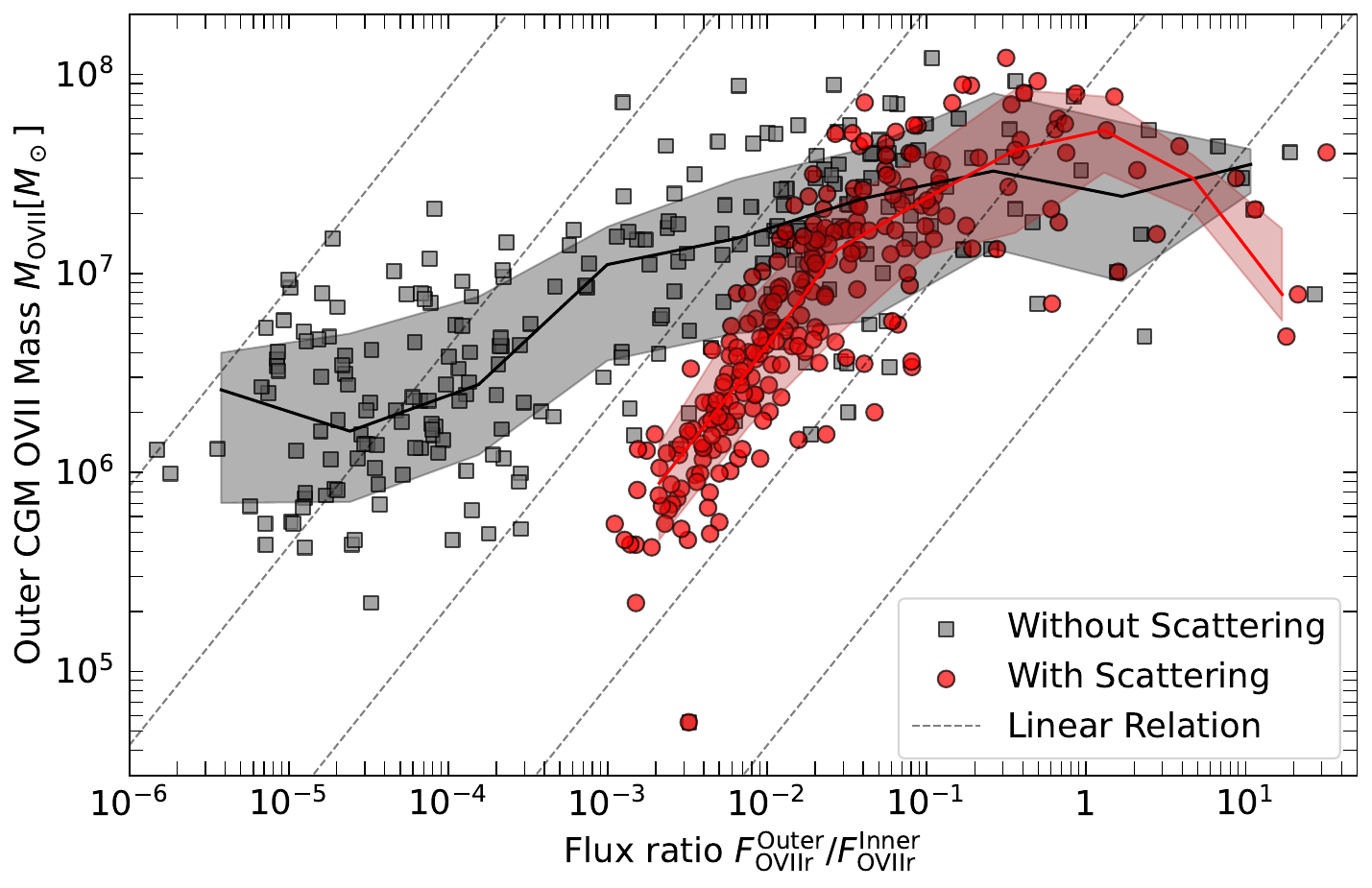}
    \caption{Effects of resonant scattering on the relation between the \ovii\ mass in the 3D $R_{500c}-R_{200c}$ shell and the \ovii{\em r}\ flux ratio between the outer and inner projected annuli. Each symbol represents a simulated galaxy in the full TNG50 sample; red and gray symbols show the flux ratios with and without resonant scattering, respectively. Shaded bands show the ${\rm 16^{th}-84^{th}}$ percentile ranges and the solid lines show the sample median. For reference, dashed lines show linear relations with different normalizations, theoretically expected from geometric arguments. With scattering included, the two quantities are tightly correlated with the expected linear slope, except at the high flux ratios. As we show below, those large deviations can be identified and excluded based on X-ray observables.} 
    \label{fig:oviir-fluxes}
    \vspace*{3mm}
\end{figure*}

\subsection{The inner and outer CGM in TNG50 galaxies}

Real galaxies do not have a well-defined boundary for the inner CGM region, where most of \ovii{\em r}\/ line emission originates. For our measurement, the inner region, which acts as an illumination source, should ideally be selected to enclose most of the \ovii{\em r}\/ photons emitted from the central region, while remaining small enough (a) to minimize scattering of the \ovii{\em r}\/ photons emitted at smaller radii, and (b) to subtend a small enough solid angle, as seen from the outer region, for the effect of any brightness asymmetries within the source region to be negligible.

The outer region should have the total observed \ovii{\em r}\/ flux dominated by the scattered photons. We select the inner and outer regions using the intrinsic and scattered surface brightness profiles for the simulated TNG50 galaxies. The radii satisfying the above criteria are straightforward to identify from these profiles.

Figure~\ref{fig:boundary} shows the cumulative profiles of the \ovii{\em r}\/ intrinsic emission for all galaxies in our simulated sample ({\it left}\/ panel), and the \ovii{\em r}\/ local surface brightness enhancement factor due to resonant scattering of the photons from the central peak ({\it right} panel). In the {\it left} panel, we show only the intrinsic \ovii{\em r}\/ emission, not including the effect of resonant scattering. We compute the radius within which $90\%$ of the intrinsic \ovii{\em r}\/ emission within $R_{\rm 500c}$ originates, $r_{90}$. For the sample of TNG50 simulated galaxies, the population median value of $r_{90}$ is about $0.2R_{\rm 500c}$. We adopt this radius as the size of the inner CGM for each individual galaxy. 
 
As shown in the {\it right}\/ panel of Figure~\ref{fig:boundary}, the median enhancement profile increases with galactocentric distance and flattens in the galaxy outskirts ($r\gtrsim R_{\rm 500c}$) around factor $\sim10$ --- most of the \ovii{\em r}\/ photons observed in the outer CGM originate from scattering rather than in-situ emission. Thus, for our experiement, we select the outer region to be between $R_{\rm 500c}-R_{\rm 200c}$ --- the radii where direct measurements of the CGM mass would be extremely difficult. Also for this selection of radii, the central region's angular size is small as seen by each element in the outer region, which minimizes the effects of any central brightness asymmetries.

We examine the mass fraction of the inner and outer CGM regions as a function of galaxy stellar mass ($M_*$). On average, the median mass fraction of the inner region is $\sim19\%$, while the outer region accounts for $\sim25\%$ of the total CGM mass measured within $R_{\rm 200c}$. However, these mass fractions display opposite trends with galaxy mass. As $M_*$ increases, the inner mass fraction decreases, whereas the outer fraction increases steeply, reaching to $\sim40\%-50\%$ for galaxies with stellar masses comparable to or exceeding that of the Milky-Way ($M_*\gtrsim5\times10^{10}M_\odot$). The latter trend may be attributed to the ejective effects of SMBH kinetic feedback in the TNG simulations, which redistributes CGM gas to larger radii in massive galaxies \citep{zinger.etal.2020, ayromlou.nelson.pillepich.2022}.

\subsection{Non-scattered vs scattered scaling relations}

With the inner and outer CGM regions defined, we can now measure the \ovii{\em r}\/ fluxes in both regions, extract the \ovii\ mass in the outer CGM of the simulated galaxies ($M_{\rm OVII}$), and examine the relationship between the flux ratio and $M_{\rm OVII}$. Figure~\ref{fig:oviir-fluxes} shows the \ovii\ mass in the $R_{\rm 500c}-R_{\rm 200c}$ 3D radial shell vs.\ the outer-to-inner flux ratio $F_{\rm OVIIr}({R_{\rm 500c}-R_{\rm 200c}})/F_{\rm OVIIr}(r<0.2R_{\rm 500c})$ in the annular sky regions (as in real observations) for our entire sample of TNG50 galaxies.

\begin{figure*}
    \centering
    \includegraphics[width=0.97\textwidth]{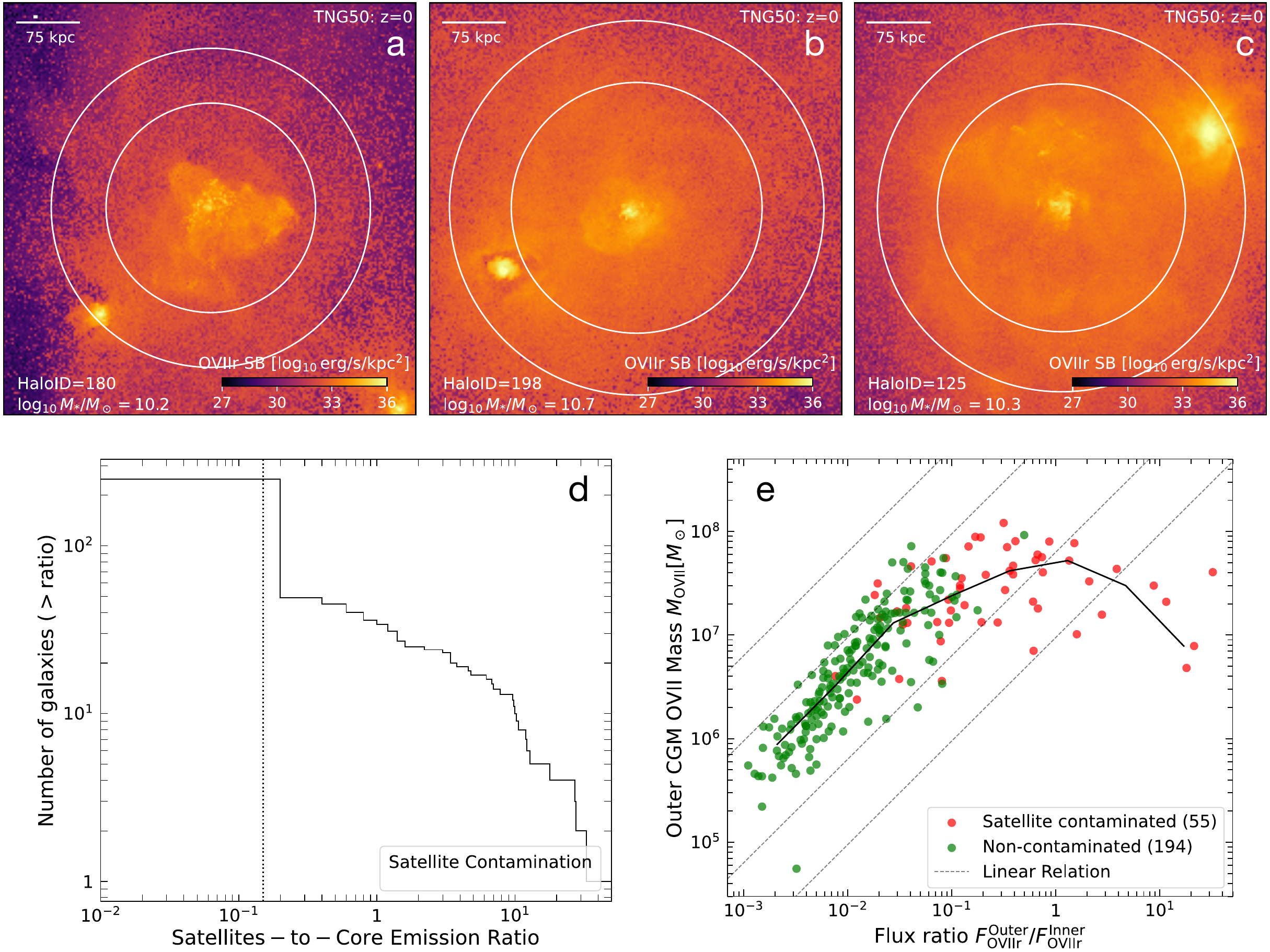}
    \vspace*{2mm}
    \caption{Impact of satellite contamination on the scattering $M_{\rm OVII}$-flux ratio relation. {\it (a), (b), and (c):} Examples of individual galaxies with satellite contamination. Shown are \ovii{\em r}\/ surface brightness maps, each 500 kpc on a side. The white circles mark the radii $R_{\rm 500c}$ and $R_{\rm 200c}$. {\it (d):} cumulative distribution of the ratio of satellite to core \ovii{\em r}\/ emission (see text for details). The vertical dotted line indicates a $15\%$ threshold. {\it (e): } The $M_{\rm OVII}$-flux ratio relation for satellite-contaminated (red) versus non-contaminated (green) galaxies. The solid line shows the running median. Dashed lines represent the theoretically-expected linear relations with different normalizations. Galaxies with significant satellite contamination systematically flatten the mass-flux ratio relation relative to the expected linear trend.} 
    \label{fig:satellite}
    \vspace*{3mm}
\end{figure*}

\begin{figure*}
    \centering
    \includegraphics[width=0.97\textwidth]{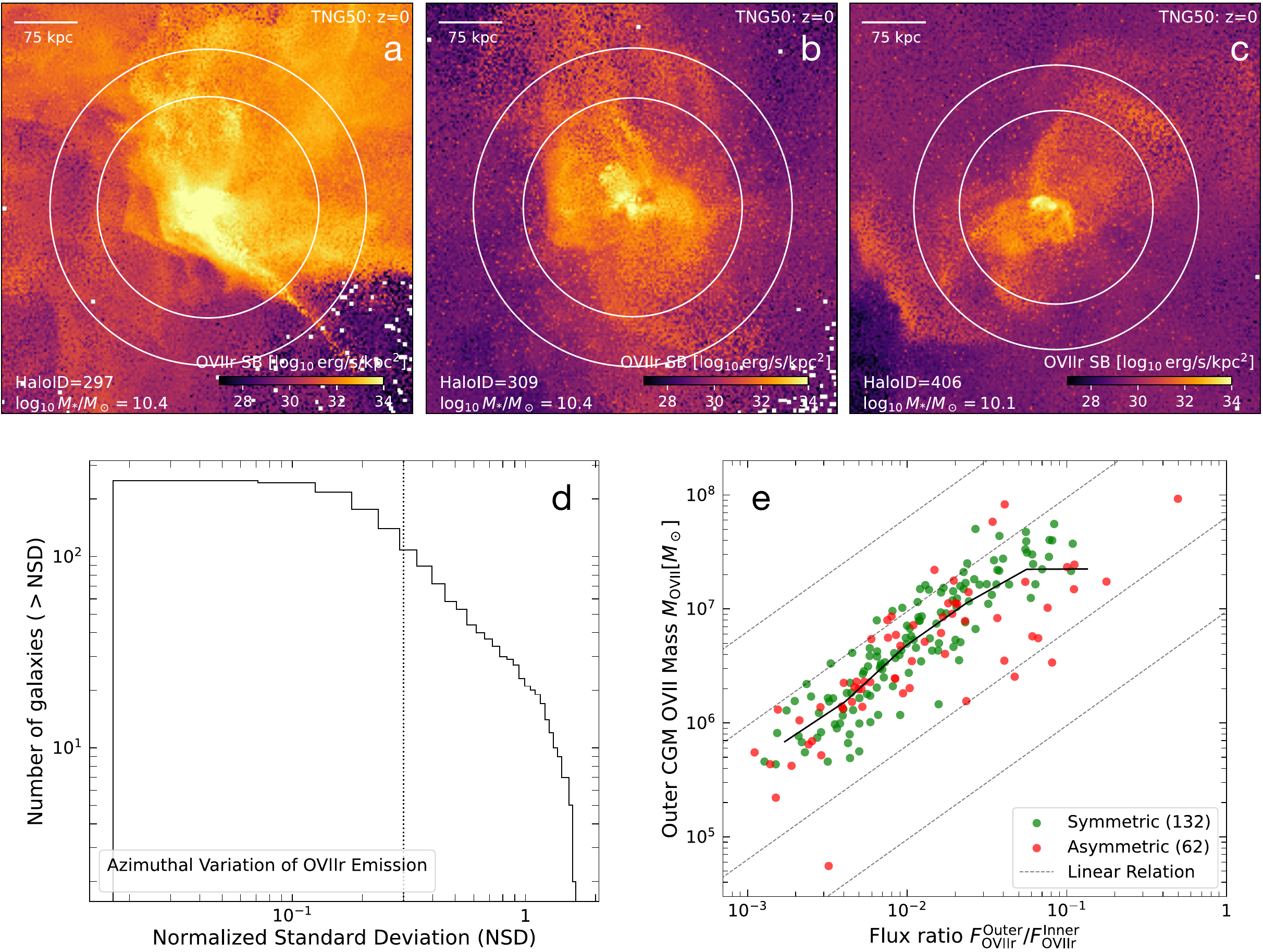}
    \vspace*{2mm}
    \caption{Impact of azimuthal anisotropy in \ovii{\em r}\/ emission on the $M_{\rm OVII}$-flux ratio relation. {\it (a), (b), and (c):} Examples of individual galaxies that exhibit significant anisotropy in their \ovii{\em r}\/ surface brightness maps. The map size and annotations follow the same conventions as Figure \ref{fig:satellite}. {\it (d):} Histogram of the standard deviation of \ovii{\em r}\/ emission across four azimuthal quadrants, normalized by the mean emission across the four quadrants. The quadrants are defined within an annulus bounded by $0.2R_{\rm 500c}$ and $R_{\rm 200c}$. The vertical dotted line indicates the threshold value of 0.3. {\it (e):} The scaling relation for both symmetric (green) and asymmetric (red) galaxies---after excluding satellite-contaminated systems---where asymmetric galaxies are defined as those with a normalized standard deviation exceeding the threshold. Despite a few outliers among the asymmetric galaxies, both subsets are broadly consistent with the expected linear relation.} 
    \label{fig:anisotropy}
    \vspace*{3mm}
\end{figure*}

We compare the cases with and without the inclusion of resonant scattering. As expected, for every galaxy, the flux ratio in the resonant scattering case is substantially higher, particularly for the low-$M_{\rm OVII}$ galaxies, where the resonant scattering enhancement in the outer annulus reaches 2--3 orders of magnitude, consistently with findings by \cite{nelson.etal.2023}.

More importantly, with scattering included, the relationship between $M_{\rm OVII}$ and the flux ratio tightens toward a linear relation, as expected from our geometric arguments (eq.\ \ref{eq:movii}), suggesting that our method for $M_{\rm OVII}$ estimation may work. However, the galaxies with the highest flux ratio deviate from the linear trend. This behavior, and how to account for it, is investigated in the next section.
 
\section{Systematic effects}
\label{sec:systematics}

In this Section, we consider how realistic galaxies deviate from the idealized picture of Sec.\ \ref{sec:method} and explore how observables, such as the X-ray surface brightness maps and emission line profiles, can be used to minimize these complications for the mass-flux ratio scaling.

\subsection{Satellite contamination}
\label{sec:satellites}

To investigate the origin of the deviation from the linear scaling at the highest flux ratios, we examine the \ovii{\em r}\/ surface brightness maps. We find that in several galaxies, the emission from the outer region is heavily contaminated by satellite galaxies. A few examples are shown in Figure~\ref{fig:satellite}{\it a,b,c}. Satellite galaxies have two effects: (i) if they project onto the outer annulus, their intrinsic \ovii{\em r}\/ emission boosts the flux from the annulus, (ii) even when they are located outside the outer annulus but close in 3D, they provide additional \ovii{\em r}\/ illumination, which is then scattered by \ovii\ ions in the outer annulus. Ignoring this contamination would lead to the behavior seen at the high-ratio end of Figure\ \ref{fig:oviir-fluxes}.

To quantify the level of this satellite contamination, for each simulated galaxy, we measure the intrinsic \ovii{\em r}\/ emission from all the satellites and compare it to the emission from the core of the main halo. We include all satellites outside the galaxy core ($r>0.2 R_{\rm 500c}$) and within twice the halo virial radius ($r<2R_{\rm 200c}$) in 3D. The satellite CGM emission is integrated within the radius that is 5 times its stellar half-mass radius. Figure~\ref{fig:satellite}{\it d} presents the distribution of the satellite-to-core emission ratio for all the studied galaxies. About 22\% of the galaxies in our sample have the contamination level above 15\%, which we will use below as a threshold to select a clean sample. Further inspection reveals that, for a given stellar mass, galaxies residing in more massive halos show higher levels of satellite contamination (not shown).

\begin{figure*}
    \centering
	\includegraphics[width=0.98\textwidth]{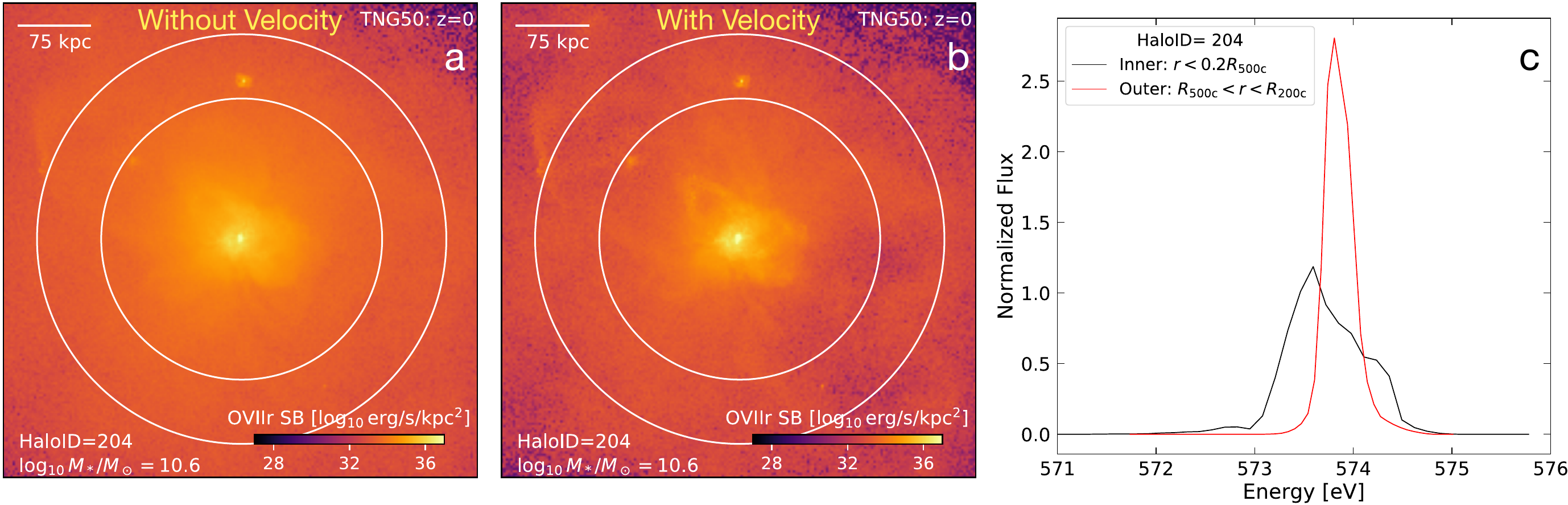}
    \vspace*{1mm}
    \caption{Effects of gas motion in the CGM on \ovii{\em r}\/ resonant scattering. {\it (a) and (b):} \ovii{\em r}\/ surface brightness maps of a TNG50 galaxy for two cases: without and with the velocity field included, respectively. The map size and annotations follow the same conventions as Figure \ref{fig:satellite}. Gas motions in the CGM suppress the scattered flux outside the core ($\gtrsim30$ kpc) by a factor of $\sim3$ relative to the velocity-free case. {\it (c):} \ovii{\em r}\/ line profiles of the same galaxy with the velocity field included, shown for the inner and outer CGM regions.} 
    \label{fig:velocity_effect1}
    \vspace*{3mm}
\end{figure*}

\begin{figure*}
    \centering
	\includegraphics[width=0.98\textwidth]{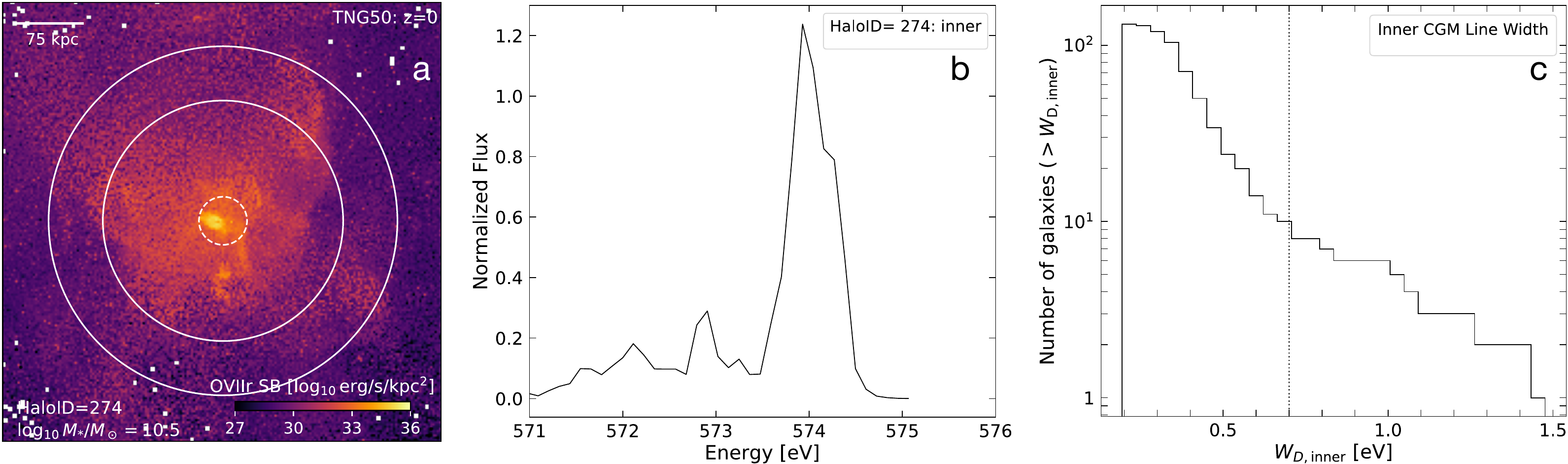}
    \vspace*{1mm}
    \caption{Effects of large-scale outflows in the inner CGM on \ovii{\em r}\/ resonant scattering. {\it (a)}: Surface brightness map of an example TNG50 galaxy exhibiting a powerful SMBH-driven outflow in the inner CGM. The map size and annotations follow the same conventions as Figure \ref{fig:satellite}, with the addition that the dashed circle marks the radius $0.2R_{\rm 500c}$, which defines the boundary of the inner CGM. {\it (b)}: \ovii{\em r}\/ line profile from the inner CGM (within the dashed circle) of this galaxy, exhibiting a broad wing ($W_{\rm D,inner}\approx1.5$ eV) attributable to the outflow. {\it (c)}: Distribution of the Doppler line widths for the inner (source) regions across the galaxy sample. Galaxies with an inner Doppler width exceeding $0.7$ eV (vertical dotted line) are deemed too strongly affected by outflows and are excluded from the analysis sample.} 
    \label{fig:velocity_effect2}
    \vspace*{3mm}
\end{figure*}

Figure~\ref{fig:satellite}{\it e} shows the mass-flux ratio relation for the galaxies above and below the 15\% contamination threshold. As expected, the contaminated galaxies (red) exhibit significantly higher flux ratios than non-contaminated galaxies (green), causing the large deviation from the linear relation. When we consider only the non-contaminated subset, the trend is in good agreement with the predicted linear behavior.

In principle, one can try to mitigate the satellite contamination by masking bright satellites directly in the X-ray surface brightness maps. However, even if the satellites themselves are masked, their O{\small VII}r photons still scatter off the CGM of the main halo and thereby contribute to the observed outer flux --- an effect that cannot be removed by simple masking without the knowledge of 3D geometry. A statistical correction based on the known satellite population from optical observations may offer a partial remedy, but would require modeling of the scattering geometry, which would increase the uncertainty of the resulting mass estimate.

\subsection{CGM Asymmetry}
\label{sec:asymmetry}

Another important systematic is the spatial anisotropy of the CGM, particularly at radii beyond the core ($r>0.2R_{\rm 500c}$). The theoretical relation in equation (\ref{eq:movii}) assumes that the CGM is isotropic, such that the scattered \ovii{\em r}\/ flux is largely independent of the projection direction. However, in practice, the CGM is disturbed by a range of dynamical processes including mergers, gas accretion, or large-scale outflows driven by feedback energy injected by the central SMBH. These disturbances can produce CGM anisotropy in terms of its thermodynamics, metal content, and gas motions \citep{peroux.etal.2020, truong.etal.2021b, nica.etal.2022}. 

In Figure \ref{fig:anisotropy}, we explore the impact of the anisotropy of temperature and metallicity on the flux ratio -- mass scaling relation. The {\it (a), (b),} and {\it (c)} panels displays examples of galaxies with significantly anisotropic \ovii{\em r}\/ surface brightness. For the galaxy in panel {\it (c)}, large-scale outflows are clearly visible, manifesting as bipolar features in the direction perpendicular to the galaxy disk \citep{nelson.etal.2019b,pillepich.etal.2021}. In others, the CGM asymmetry is likely driven by mergers and accretion.

\begin{figure*}[ht!]
    \centering
    \includegraphics[width=0.97\textwidth]{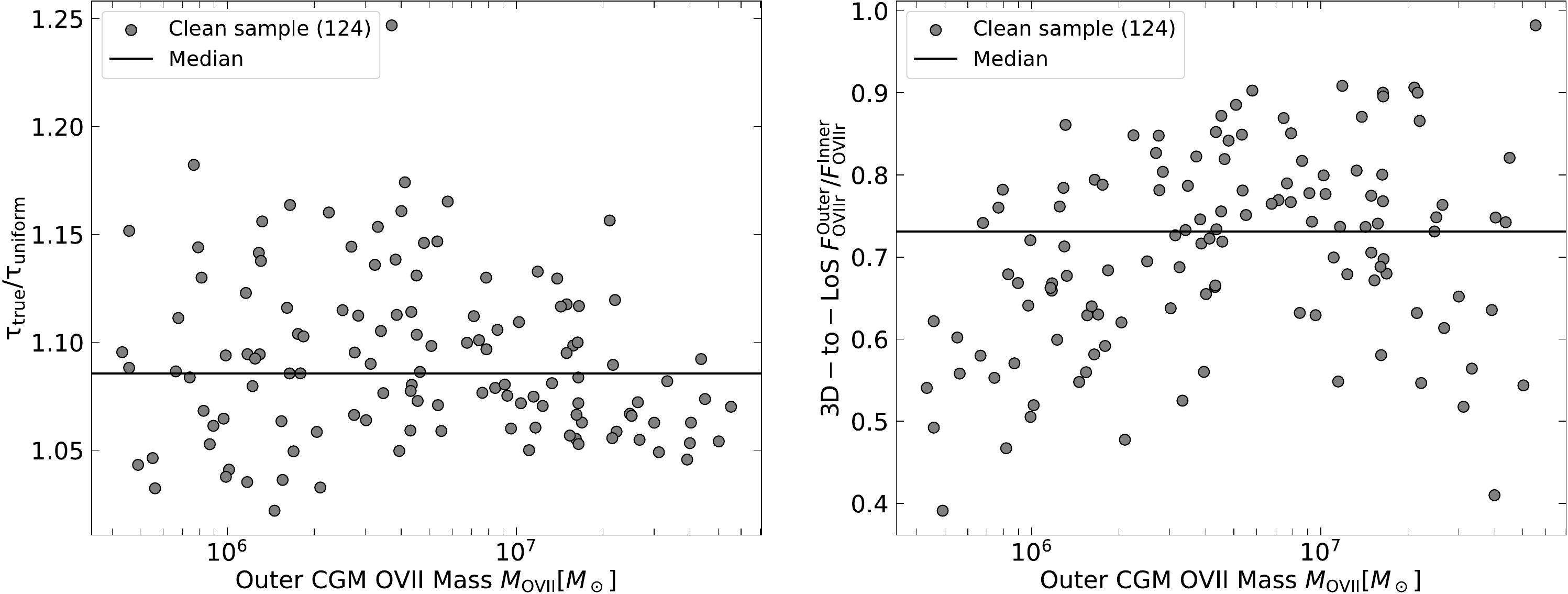}
    \vspace*{1mm}
    \caption{Effects of shell thickness and projection. {\it Left: }Ratio of true optical depth to that of a uniform density shell as a function of the outer CGM \ovii\ mass. {\it Right: }  Ratio of the scattered flux ratio $F_{\rm OVIIr}^{\rm Outer}/F_{\rm OVIIr}^{\rm Inner}$ computed using only emission within the 3D inner and outer regions to that computed using all the line-of-sight emission.} 
    \label{fig:projection_thickness}
    \vspace*{3mm}
\end{figure*}

To quantify the degree of anisotropy, we calculate the standard deviation in the \ovii{\em r}\/ surface brightness across four 90$^\circ$  quadrants within a broad annulus between $0.2R_{\rm 500c}<r<R_{\rm 200c}$. Figure \ref{fig:anisotropy}{\it d} presents the distribution of this standard deviation normalized by the mean surface brightness of the galaxy, NSD, across the galaxy sample. We adopt a fiducial threshold of ${\rm NSD\geq0.3}$ to classify galaxies as anisotropic. About $32\%$ of the satellite-excluded galaxies lies above this threshold. In Figure \ref{fig:anisotropy}{\it e}, we examine where these anisotropic galaxies fall in the scaling relation, from which the satellite-contaminated galaxies have already been excluded (Sec.\ \ref{sec:satellites}).

The adopted threshold excludes the biggest outliers (red), whose flux ratios are significantly higher or lower than the median at a given $M_{\rm OVII}$. However, it also excludes a number of galaxies on the median of the relation. This may arise because the anisotropy is assessed using 2D surface brightness maps (as the observer would be able to do), rather than the 3D distribution of CGM gas properties. The trends between the symmetric and asymmetric subsets are similar and largely consistent with the expected linear relation, but the symmetric subset has a lower scatter.

Correcting for CGM asymmetry rather than excluding the most asymmetric galaxies would require the knowledge of the 3D gas distribution, without which the scatter of the mass estimate would inevitably increase.

\subsection{Impact of CGM gas motions}

\subsubsection{Gas turbulent and bulk motions}

Gas motions in the CGM reduce the scattered signal through two distinct effects: Doppler broadening and Doppler shifting, as described in equation~(\ref{eqn5}). Doppler broadening is the sum of thermal ion motions and kinematics of the CGM (equation~\ref{eqn6}) and reduces the peak cross-section at the line center. In addition to random motions, coherent bulk flows can further suppress resonant scattering by Doppler-shifting the energy of the incident \ovii{\em r}\/ photons in the rest frame of the scattering ions away from the line center, where the cross-section is maximal, thereby reducing the scattering efficiency or preventing the scattering completely if the shift is sufficiently large. This is the most significant systematics affecting the accuracy of our method.

To explore the magnitude of this effect, we re-run the scattering RT calculations for selected galaxies and intentionally neglect gas velocities, assuming a static CGM and including only the thermal broadening of the spectral lines. Figure \ref{fig:velocity_effect1}{\it a} and {\it b} compares the \ovii{\em r}\/ surface brightness maps of a TNG50 MW-like galaxy ($M_*\sim10^{10.7}\rm{M}_\odot$) in two cases with the velocity field disable and enabled, respectively. When gas motions are included, the \ovii{\em r}\/ surface brightness decreases significantly across much of the CGM (approximately by a factor of 3), except in the central region where the intrinsic emission dominates. This underscores the need to account for this effect when attempting to use scattering for the mass estimate.

We address this by using the \ovii{\em r}\/ line profiles from the inner and outer CGM regions, both of which will be directly observable in a future experiment. The example line profiles, with the velocity field included, are shown in Figure~\ref{fig:velocity_effect1}{\it c}. The outer line profile is as observed from our line of sight, i.e., it includes the effect of the gas velocities on scattering. Under the assumption of isotropic random gas velocities in the inner (source) and outer (scatterer) regions, and the expectation that the velocity distribution in the outer shell is not broader than that in the source region (and so we are not missing a large amount of gas at the wings of the velocity distribution that is not scattering because of the Doppler mismatch), we can rewrite eq.\ (\ref{eqn:sigma_scat}) for the averaged cross-section using the {\em line-of-sight, observed}\/ energies and widths as:
\begin{equation}
    \langle\sigma_{\rm scat}\rangle\approx \frac{\sqrt{\pi} h r_e c f}{\sqrt{W_{D, {\rm inner}}^2+W_{D, {\rm outer}}^2}}\exp{\bigg[-\frac{\big(E_{\rm inner}-E_{\rm outer}\big)^2}{W_{D, {\rm inner}}^2+W_{D, {\rm outer}}^2}\bigg]},
\label{eqn:averaged_cross_section} 
\end{equation}
where $E_{\rm inner}$ and $E_{\rm outer}$ are the line centroids of the inner and outer regions, respectively, and $W_{D, {\rm inner}}$ and $W_{D, {\rm outer}}$ are their {\em observed}\/ Doppler widths.

\subsubsection{SMBH-driven large-scale outflows}
%%%%%%%%%%%%%%%%%%%%%%%%%%%%%%%%%%%%%%%%%%%%%%%%%%%%%%%%%%%

It is important to note that the measured profile broadening is not always caused by small-scale turbulent motions. Some galaxies exhibit large-scale outflows driven by SMBH feedback. Those outflows reshape the CGM and make it less isotropic, as discussed above, but also significantly impact velocity space as they progress outwards mainly along the minor axis directions of galaxies i.e. with broadly bipolar-like geometry \citep{nelson.etal.2019b, pillepich.etal.2021,truong.etal.2021b}.

We assess the impacts of large-scale outflows in the bottom panels of Figure \ref{fig:velocity_effect2}. In Figure \ref{fig:velocity_effect2}{\it a}, we show the surface brightness map of an example galaxy with a prominent outflow. Outflow velocities can easily reach several hundred km\,s$^{-1}$, causing significant \ovii{\em r}\/ line profile distortion, as shown in Figure \ref{fig:velocity_effect2}{\it b}. Since outflow motions are directional rather than random, this breaks the above assumption, under which we estimate the illuminating line width as seen by the scattering ions from the line width observed along our line of sight 
 --- the directions that are mostly perpendicular to each other.
Since the outflow angle with respect to the line of sight is not known, 
it is best to exclude the galaxies obviously affected by such outflows from our mass test. We do it 
based on the inner-region line width. Figure \ref{fig:velocity_effect2}{\it c}\/ represents the distribution of the inner width across the sample. We adopt a threshold of $W_{\rm D, inner}>0.7$ eV to exclude the galaxies impacted by outflows, which removes $\sim6\%$ of the satellite-excluded, symmetric sample. 

Our final sample of simulated galaxies excludes those affected by strong satellite contamination, CGM anisotropy, and outflows. The selection for each of these effects is based on {\em observable}\/ quantities and can be done in a future experiment without reliance on simulations. This ``clean sample'' consists of 124 galaxies, $\sim 50\%$ of the original full set. We will evaluate the accuracy of the \ovii\ mass derivation for this clean sample based on the theoretical relation (equation~\ref{eq:movii}) in Section~\ref{sec:theory_vs_sim}.

%%%%%%%%%%%%%%%%%%%%%%%%%%%%%%%%%%%%%%%%%%%%%%%%%%%
\subsection{Shell thickness and projection effects}
\label{sec:thickness_projection}

The idealized case assumes that the gas density within the outer shell is uniform. In realistic CGM, the gas density usually declines with radius following a $\beta$-model-like profile. To evaluate the geometric effect of shell thickness and non-uniform density on the resonant scattering process, we compare the optical depth, which is defined as $\tau=\int_{R}^{R+\Delta R} \rho_{\rm gas}dr$, computed from the simulated CGM density profiles with that of an equivalent uniform-density shell containing the same total gas mass. The {\it left} panel of Figure~\ref{fig:projection_thickness} shows the ratio of these two optical depths as a function of the outer ($R_{500c}-R_{200c}$ shell) CGM \ovii\ mass for the clean sample. Real systems exhibit systematically higher optical depths than their uniform-shell counterparts, by $8\%$ on average. This is expected because a more centrally concentrated \ovii\ distribution places a greater fraction of ions at smaller radii, where they intercept a greater fraction of the \ovii{\em r}\/ photons from the center compared to the uniform-density case.

Another observational effect that must be accounted for is projection. In the idealized case, all scattered \ovii{\em r}\/ photons originate in the 3D outer shell. In the real experiment using the galaxy image, (i) only a fraction of the shell volume projects onto the corresponding annular region in the sky; and (ii) the annulus also contains photons scattered at larger radii on the line of sight. The first effect can be addressed through simple geometric computation of the intersection between the shell and the cylinder defining the annulus in the sky. This provides a correction for the incomplete coverage of the shell in projection. 

The second effect can be corrected by evaluating the emission from the gas along the line of sight beyond the 3D shell. Here, we estimate this contribution directly from the simulated three-dimensional \ovii{\em r}\/ emission. The {\it right} panel of Figure~\ref{fig:projection_thickness} shows the ratio of the scattered flux ratio, $F_{\rm OVIIr}^{\rm outer}/F_{\rm OVIIr}^{\rm inner}$, computed using only the emission within the 3D inner and outer shells, to the same ratio but computed using all emission along the light-of-sight. We use  $r<0.2R_{\rm 500c}$ for the inner region and $R_{\rm 500c}<r<R_{\rm 200c}$ for the outer region. On average, the flux ratio computed from the 3D regions is approximately $73\%$ of that computed from the full line-of-sight integration. 

We apply these average correction factors, which account for shell thickness and projection effects for our particular choice of the shell radii, to the predicted idealized scaling relation (equation~\ref{eq:movii}). Note that these trivial corrections should be largely insensitive to the simulation details and could instead be derived with sufficient accuracy by, e.g., fitting and extrapolating a $\beta$-model profile for the observed \ovii\ line surface brightness.

\begin{figure}[ht!]
    \centering
    \includegraphics[width=0.46\textwidth]{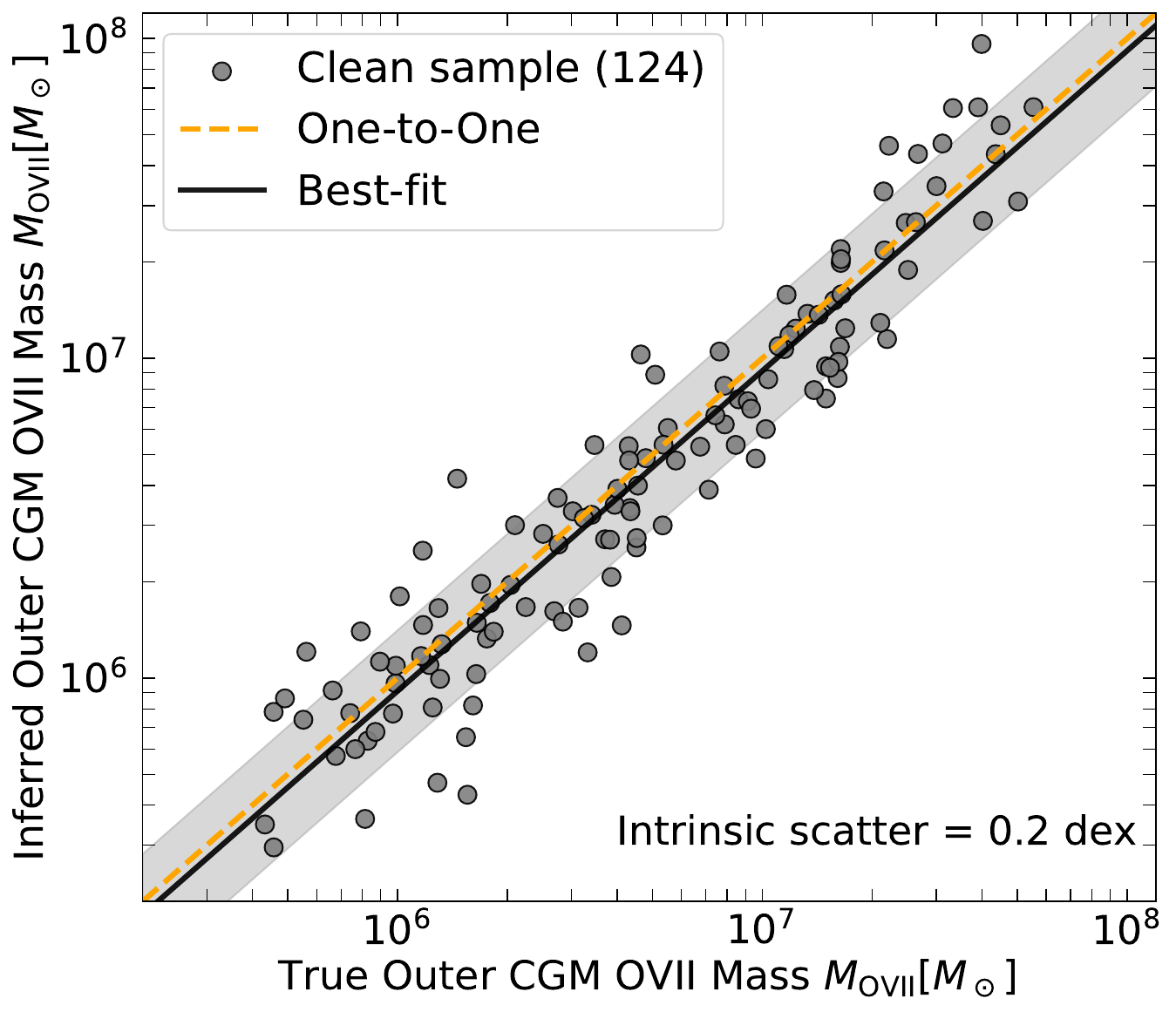}
    \caption{Comparison between the \ovii\ mass in the $R_{\rm 500c}-R_{\rm 200c}$ 3D shell inferred from the scaling relation (equation \ref{eqn:MOVII_Flux}) and the true \ovii\ mass extracted directly from the simulation, for each of the 124 galaxies in the clean sample (Section~\ref{sec:systematics}). The shaded area indicates the $16^{\rm th}-84^{\rm th}$ percentile scatter around the best-fit relation. The inferred and true \ovii\ masses follow a near one-to-one relation across the full considered mass range, with an intrinsic rms scatter of $\sim0.2$ dex. The inferred masses are systematically underestimated by $\sim10\%$ relative to the true values.} 
    \label{fig:predicted_vs_sim}
    \vspace*{2mm}
\end{figure}
\section{Predicted vs. simulated masses}
\label{sec:theory_vs_sim}
%%%%%%%%%%%%%%%%%%%%%%%%%%%%%%%%%%%%%%%%%%%%%%%%%%%

We now compare the \ovii\ mass estimates from our geometric method (equation\ \ref{eq:movii}) with the \ovii\ masses extracted directly from the TNG50 simulation, using the clean galaxy sample and incorporating the systematic corrections from Section~\ref{sec:systematics}. Based on equation~(\ref{eq:movii}), the \ovii\ mass in the $R_{\rm 500c}-R_{\rm 200c}$ shell can be expressed as
\begin{equation}
    M_{\rm OVII} = A_{O}m_p\frac{ 4\pi \big(R_{\rm 500c}^2+R_{\rm 500c}R_{\rm 200c}+R_{\rm 200c}^2\big)}{3\langle\sigma_{\rm scat}\rangle}C_{\tau}C_{\rm proj}\bigg(\frac{F_{\rm OVIIr}^{\rm Outer}}{F_{\rm OVIIr}^{\rm Inner}}\bigg), 
\label{eqn:MOVII_Flux}
\end{equation}
where $C_\tau\approx0.92$ and
\begin{equation}
C_{\rm proj}\approx 0.73 \frac{R_{\rm 500c}^2+R_{\rm 500c}R_{\rm 200c}+R_{\rm 200c}^2}{(R_{\rm 200c}+R_{\rm 500c})^{3/2}(R_{\rm 200c}-R_{\rm 500c})^{1/2}} \nonumber
\end{equation}
are correction factors accounting for the shell thickness and projection effects, respectively, as discussed in Section~\ref{sec:thickness_projection}. The numerical prefactors of these corrections are based on simulation-informed average values shown in Figure~\ref{fig:projection_thickness}. For the projection correction, $C_{\rm proj}$, we include an explicit geometric factor determined by the outer shell's boundaries, i.e. $R_{\rm 500c}$ and $R_{\rm 200c}$. The averaged scattering cross-section, $\langle\sigma_{\rm scat}\rangle$, is estimated using equation~(\ref{eqn:averaged_cross_section}), which incorporates observable line centroids and widths from both the inner and outer CGM regions to correct for the effects of gas motions. 

We use only the scattered flux from the outer CGM region, i.e., not including its intrinsic X-ray emission. The scattered flux is dominant there for most of the sample (Fig.\ \ref{fig:boundary}). Where the intrinsic contribution is non-negligible, the two contributions could be separated using the relative strengths of the \ovii\ triplet components (\S\ref{sec:theory}). We disregard the uncertainty of such spectral separation here.

Figure~\ref{fig:predicted_vs_sim} compares the \ovii\ mass inferred from equation~(\ref{eqn:MOVII_Flux}) with the true \ovii\ mass directly measured from the simulation for the clean sample of 124 TNG50 galaxies. The inferred and true \ovii\ masses exhibit a remarkably strong correlation, following a near one-to-one relation with an intrinsic scatter of $\sim 0.2$ dex. However, there is a small systematic bias: the inferred masses are, on average, lower than the true \ovii\ masses by factor $\approx0.9$ across the sample. 

\begin{figure}
    \centering
    \includegraphics[width=0.46\textwidth]{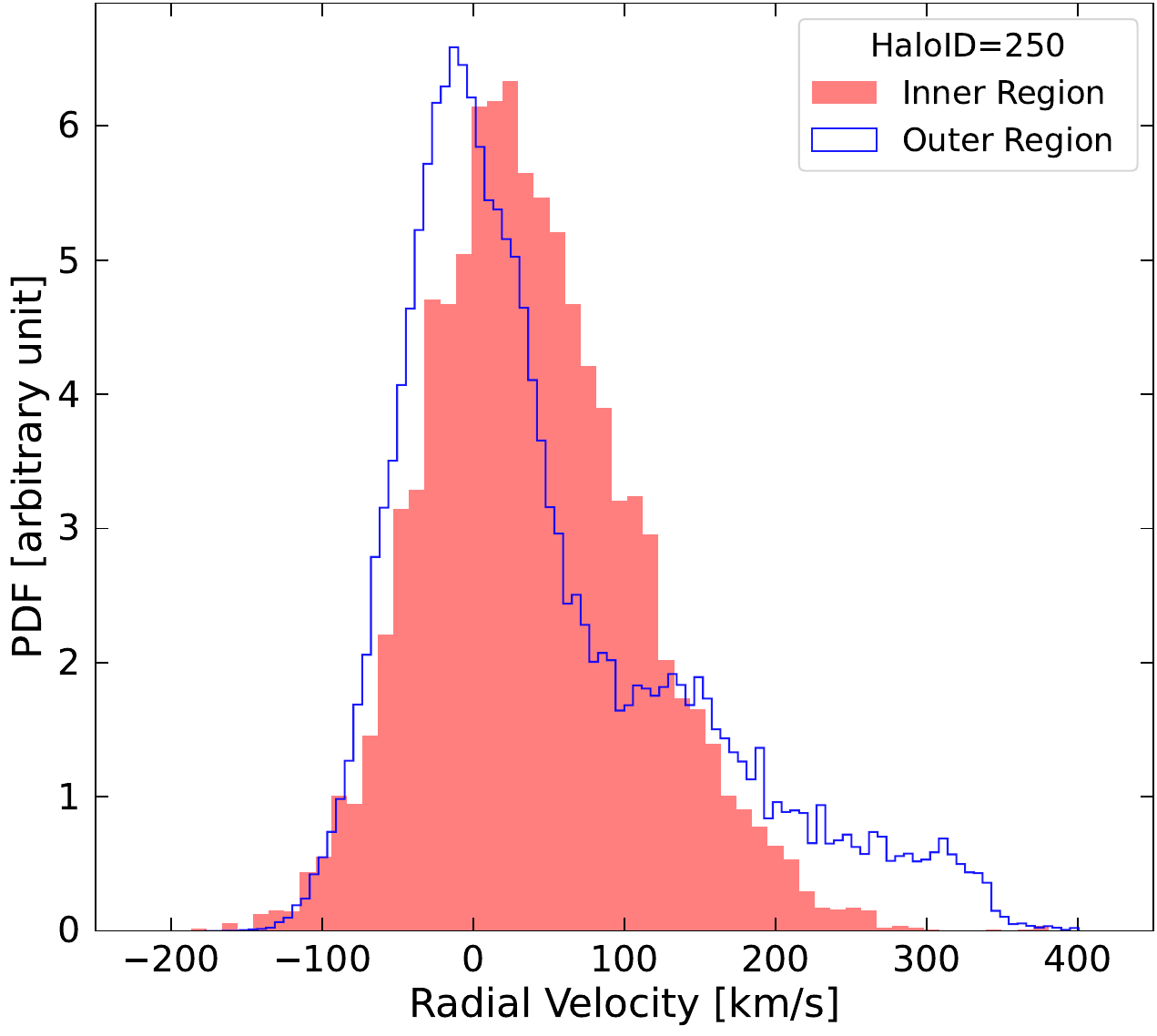}
    \caption{Distributions of radial velocities in the inner and outer CGM regions for a TNG50 galaxy. The outer CGM often exhibits high-velocity wings extending beyond the velocity range of the inner region. These high-velocity gas parcels have suppressed \ovii {\em r} scattering cross-sections due to a large Doppler mismatch, and would therefore be not captured by our mass estimate.} 
    \label{fig:vrad_dis}
    \vspace*{2mm}
\end{figure}

The residual bias indicates the presence of additional systematic effects not captured in equation~(\ref{eqn:MOVII_Flux}). It is likely caused by the presence of fast-moving gas in the outer CGM. A key assumption underlying our method is that all \ovii\ ions in the outer CGM region can scatter \ovii{\em r}\/ photons from the central source. This assumption holds only if the velocity distribution in the outer CGM overlaps sufficiently with that of the inner CGM.
Figure~\ref{fig:vrad_dis} shows the radial velocity distributions from inner and outer regions for a typical TNG50 galaxy. A fraction of the gas cells in the outer shell exhibit high velocities, extending beyond the velocity range of the inner CGM. These high-velocity \ovii\ ions would not scatter the \ovii{\em r}\/ emission from the center because of the Doppler mismatch between the wavelengths of the emitter and the scatterer, and would not be counted by our test. Unlike the systematics considered above, this effect cannot be corrected based on observables, as \ovii\ ions with velocities outside the resonant scattering window do not contribute to the observed \ovii{\em r}\/ signal.%
\footnote{With a sensitive enough instrument, the very faint {\em intrinsic}\/ X-ray emission from that fast gas phase could be detected, and at least the presence of the bias deduced from the O{\sc vii}\ triplet line ratio being close to that for the thermal emission. Unlike the scattered emission proportional to the O{\sc vii}\ density, the intrinsic X-ray emission is proportional to $\rho_{\rm gas}^2$ and, even if detected in the clumpy outskirts of a CGM halo, would not provide a gas mass estimate with any useful accuracy.}
Cosmological simulations, such as TNG50 used in this study, may be used to quantify and calibrate this bias. However, such calibration would be model-dependent, because different galaxy formation and feedback models predict different CGM velocity fields and radial CGM density profiles.
We therefore consider this bias as a measure of the systematic uncertainty of our geometry-based \ovii\ mass estimate.

\begin{figure*}[ht!]
    \centering
    \includegraphics[width=0.97\textwidth]{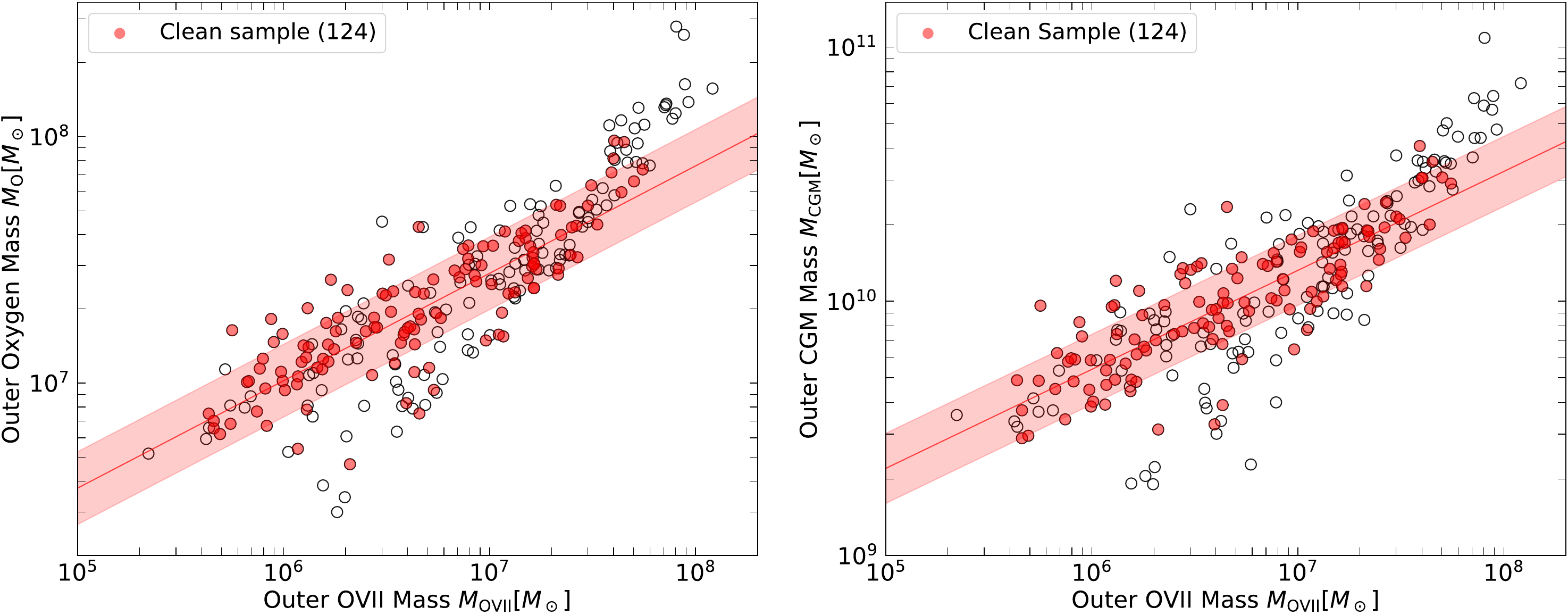}

    \caption{Mass scaling relations in the outer CGM region: \ovii\ mass vs.\ total oxygen mass ({\it left}) and CGM mass ({\it right}) in the outer $R_{\rm 500c}-R_{200c}$ shell, for all TNG50 galaxies and the clean sample (filled). Solid lines represent the best-fit scaling relations for the clean sample, while shaded bands show the $16^{\rm th}-84^{\rm th}$ percentile scatter. The clean sample exhibits strong correlations between $M_{\rm O}$ and $M_{\rm OVII}$, as well as between $M_{\rm CGM}$ and $M_{\rm OVII}$, with an intrinsic scatter of $\sim 0.15$ dex, offering a potential means to probe the outer CGM mass content.} 
    \label{fig:scalings}
    \vspace*{3mm}
\end{figure*}

We must recall here that our \ovii\ mass estimate is made under the assumption of isotropic scattering (\S\ref{sec:RTsim}). This approximation does not enter in the 10\% bias discussed above, because the simulation and the estimate both use it self-consistently. But it will result in a bias (likely $>10$\%) when our estimate is applied to real galaxies, given that it uses photons scattered predominantly at $\theta\sim90^\circ$. Therefore, before any real-world application, the calculation needs to be redone with the real-world \ovii\ scattering phase function. Since this is a purely geometric effect, it should be easy to include in the simulations and be independent of any galaxy physics. However, since all the other systematics in the method are limited to 10\%, it would be best to use the phase function measured in the laboratory rather than relying on its theoretical approximation. This is left for future work.

\subsection{Total oxygen and CGM mass estimates}
\label{sec:mass_scaling}

Our \ovii\ mass estimate provides a potential method for probing the baryonic mass content of the outer CGM, relating the \ovii\ mass with other quantities using cosmological simulations. In Figure~\ref{fig:scalings}, we show scaling relations that link different mass components in the outer CGM of TNG50 galaxies: total oxygen mass ($M_{\rm O}$), \ovii\ mass ($M_{\rm OVII}$), and total CGM mass ($M_{\rm CGM}$).

The masses of oxygen and CGM, derived from the multiphase gas ($T\gtrsim10^{4}$ K) and measured within the outer 3D annulus $R_{\rm 500c}<r<R_{\rm200c}$, exhibit strong correlations with the \ovii\ mass, particularly for the clean sample used above for the accurate \ovii\ mass. For this sample, the $M_{\rm O}-M_{\rm OVII}$ and $M_{\rm CGM}-M_{\rm OVII}$ scaling relations have an intrinsic scatter of approximately $0.15$ dex. The best-fit scaling relations are given by 
\begin{equation}
\log_{10}\frac{M_{\rm O}}{M_\odot} = 4.40+0.44\log_{10}\frac{M_{\rm OVII}}{M_\odot}\pm0.15\ {\rm dex}, 
\label{eqn:mO_flux}
\end{equation}
\begin{equation}
\log_{10}\frac{M_{\rm CGM}}{M_\odot} = 7.40+0.39\log_{10}\frac{M_{\rm OVII}}{M_\odot}\pm0.14\ {\rm dex}, 
\label{eqn:mcgm_flux}
\end{equation}
where the normalization, slope, and intrinsic scatter are treated as free parameters and fitted simultaneously by performing linear regression in logarithmic space using the \texttt{linmix\underline{ }err} package \citep{kelly.2007}. An investigation into the origin of these scaling relations is beyond the scope of this paper, it is worth noting that in TNG50, different components of the CGM mass content in the outer region, including $M_{\rm O}$, $M_{\rm OVII}$, and $M_{\rm CGM}$, as well as the CGM's average temperature, are strongly correlated with the mass of the host halo. This underlying connection may explain the correlations between $M_{\rm O}$, $M_{\rm OVII}$, and $M_{\rm CGM}$. The selection of the clean sample, which excludes satellite-contaminated and irregular CGM galaxies, likely contributes to reducing the scatter in these mass scaling relations. 

%\begin{figure}[ht!]
\begin{figure}[ht!]
    \centering
	\includegraphics[width=0.46\textwidth]{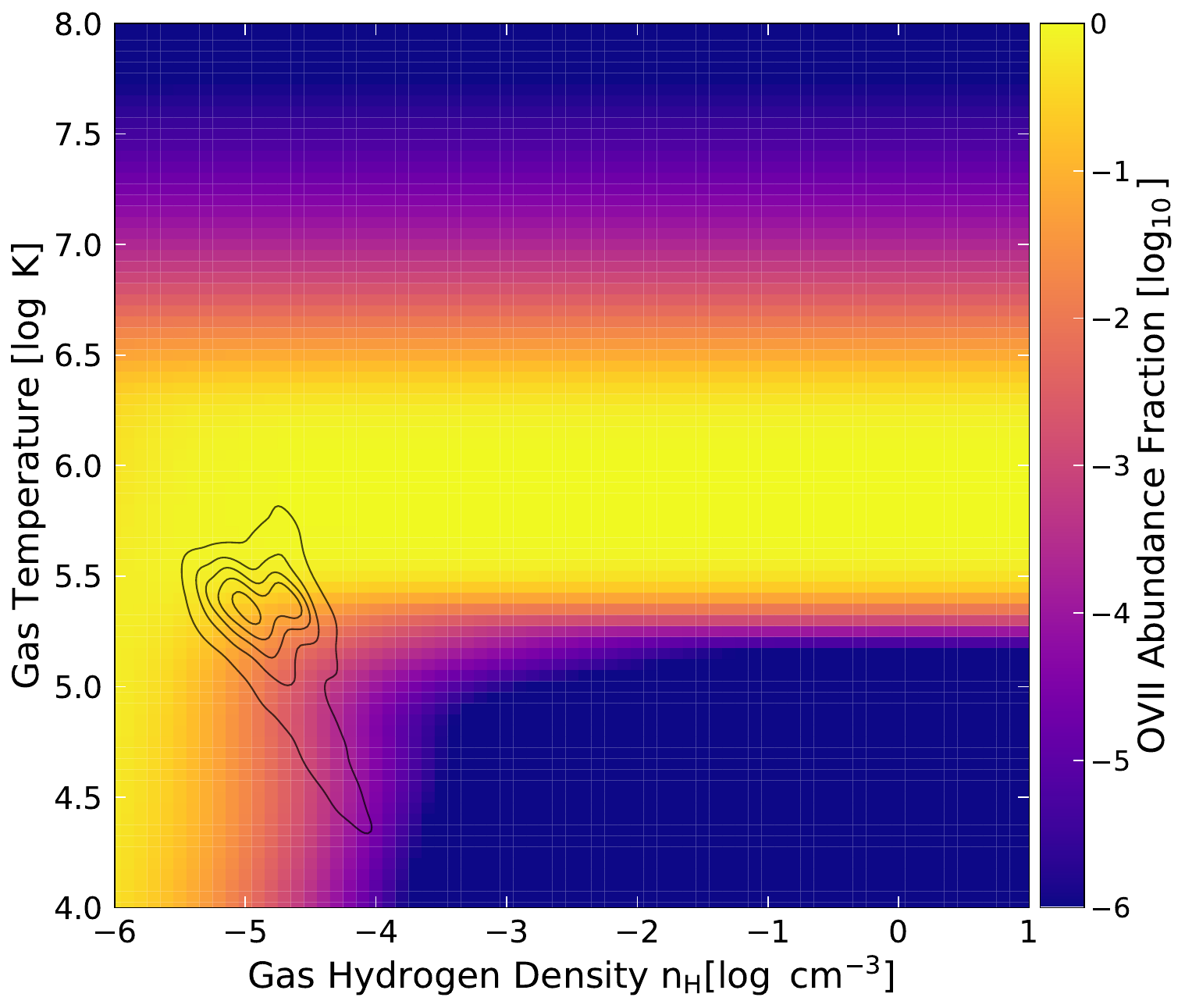}
    \caption{Phase-space diagram showing the \ovii\ fraction of the total Oxygen content (color) for different gas temperatures and densities. The \ovii\ fraction is calculated using the CLOUDY model described in Section \ref{sec:method}, which includes collisional ionization and photoionization by CXB. Contours show the typical gas parameters in our outer CGM shell, for a MW-like galaxy in TNG50. The density there is low ($n_{\rm H}\lesssim10^{-5}{\rm cm^{-3}}$) and the \ovii\ fraction is driven mostly by CXB photoionization.} 
    \label{fig:temp_rho}
    \vspace*{-2mm}
\end{figure}

It is also important to examine the Oxygen-to-\ovii\ conversion in the outer CGM. Figure \ref{fig:temp_rho} presents a phase-space diagram showing the \ovii\ abundance as a function of gas temperature and density. The abundance is calculated using the CLOUDY model described in Section \ref{sec:method}, which includes both collisional and photoionization processes. For a Milky-Way like galaxy at z=0 in TNG50, the CGM typically occupies a low-density regime in the phase-space diagram ($n_{\rm H}\lesssim10^{-5}{\rm cm^{-3}}$), where \ovii\ ionization is primarily driven by photoionization processes from the UV+X-ray background and is only weakly dependent on the temperature.

%%%%%%%%%%%%%%%%%%%%%%%%%%%%%%%%%%%%%%%

\section{Summary}
\label{sec:summary}

We propose a method to count the total number of \ovii\ ions in the outskirts of the galactic gas halos at radii where the CGM thermal X-ray emission is very faint and the usual CGM mass estimates are not feasible. The method uses resonant scattering of the \ovii{\em r}\/ photons emitted by the central, bright regions of the galactic halo, by \ovii\ ions in the outskirts. For a spherically symmetric, optically thin CGM (which is expected to be the case outside the small central regions containing the galactic disk), simple geometric considerations predict that the total number of \ovii\ ions in a given radial shell in the CGM outskirts is directly proportional to the ratio of the observed \ovii{\em r}\/ fluxes from the corresponding sky annulus and the central region, with the proportionality coefficient given by the scattering cross-section and geometric arguments.

We use a sample of galaxies with stellar masses in the range $10^{10-11}M_\odot$ from the TNG50 cosmological simulation suite, supplemented by radiative transfer modeling, to evaluate the accuracy of this method for realistic galaxies that include gas motions, satellites, asymmetries, etc., and to devise selection criteria to pick a galaxy sample for which this method would have interesting accuracy. 

We find that for a radial shell $R_{500c}<r<R_{200c}$, where the \ovii{\em r}\/ resonant scattering emission is much higher than the intrinsic thermal emission, we can deduce the true number of \ovii\ ions from the ratio of the \ovii{\em r}\/ fluxes in the $R_{500c}-R_{200c}$ annulus and the central $r<0.2R_{500c}$ region with interesting accuracy --- with a systematic bias of only 10\% and an rms scatter of 0.2 dex (Fig.\ \ref{fig:predicted_vs_sim}) across the full mass range. This accuracy is achieved for a clean galaxy sample that comprises 50\% of the total galaxy set, excluding the galaxies strongly contaminated by satellites, exhibiting extreme asymmetries or high-velocity gas outflows in the core, by applying the selection based purely on X-ray observables. The factor 0.9 bias should be caused by the presence of high gas velocities in the CGM outskirts, which makes a fraction of the outer \ovii\ ions invisible to this measurement because of the Doppler mismatch of the \ovii{\em r}\/ wavelengths. 

The \ovii\ mass can be used to deduce the total oxygen and CGM baryonic mass using the correlations between these quantities predicted by the simulations. These correlations are especially tight for the clean galaxy sample, with a relatively small scatter of $\sim0.15$ dex.

It will be important to perform a similar evaluation using cosmological simulations that employ different prescriptions for galactic feedback, which may generate different gas velocities in the outskirts and therefore a different bias and scatter for the method. The ultimate purpose of measuring the oxygen and total gas masses in the CGM is to constrain models of galaxy formation.

Our geometry-based \ovii\ ion counting method can be applied to CGM mapping observations with the future X-ray microcalorimeter missions that will be capable of spectrally resolving the \ovii\ emission line triplet components and separating them from the Milky Way foreground, such as {\em NewAthena} and {\em HUBS}.
We would apply the X-ray selection criteria similar to those we used above to a dataset of galaxies observed with such an instrument. The selected galaxy sample would yield the oxygen and CGM mass estimates with the accuracy that may allow us to discriminate between different models of galaxy formation.

\begin{acknowledgements}
The idea of this method was developed during the work on the {\em LEM}\/ Astrophysics Probe concept proposed to NASA in 2023. This research was supported by NASA under award number 80GSFC24M0006. DN acknowledges funding from the Deutsche Forschungsgemeinschaft (DFG)
through an Emmy Noether Research Group (grant number NE 2441/1-1).
\end{acknowledgements}

\bibliography{refs}{}
\bibliographystyle{aasjournal}
\end{document}